\documentclass[12pt]{article}

\def\tr{{\rm tr}}
\def\Tr{{\rm Tr}}

\usepackage{fullpage,epsfig,fancyheadings}

\usepackage[psamsfonts]{euscript}

\usepackage{epic}

\usepackage{amstex}

\newcommand{\beq}{\begin{equation}}
\newcommand{\eeq}{\end{equation}}

\begin{document}
\vspace*{-.6in}
\thispagestyle{empty}

\begin{flushright}
CALT-68-2030\\
hep-th/9512053
\end{flushright}
\baselineskip = 20pt

\vspace{.5in}

{\Large
\begin{center}
Anomaly-Free Supersymmetric Models in Six Dimensions\footnote{Work
supported in part by
the U.S. Dept. of Energy under Grant No. DE-FG03-92-ER40701.}
\end{center}}

\vspace{.4in}

\begin{center}
John H. Schwarz\\
\emph{California Institute of Technology, Pasadena, CA  91125, USA}
\end{center}
\vspace{1in}

\begin{center}

\textbf{Abstract}

\end{center}

\begin{quotation}
\noindent The conditions for the cancellation of all gauge,
gravitational, and mixed anomalies of $N=1$ supersymmetric models
in six dimensions are reviewed and illustrated by a number of examples.
Of particular interest are models that cannot be realized perturbatively
in string theory. An example of this type, which we verify satisfies
the anomaly cancellation conditions, is the K3 compactification
of the $SO(32)$ theory with small instantons recently proposed by Witten.
When the instantons coincide it has gauge group $SO(32) \times Sp(24)$.
Two new classes of models, for which non-perturbative string constructions
are not yet known, are also presented. They have gauge groups
$SO(2n+8)\times Sp(n)$ and $SU(n)\times SU(n)$, where $n$ is an
arbitrary positive integer.

\end{quotation}
\vfil

\newpage
\pagenumbering{arabic}

\section{Introduction}
Recent developments have made it clear that the various known
``superstring theories'' and their compactifications are actually
recipes for constructing solutions to a unique underlying theory
\cite{sen}--\cite{horava}.
Even though this theory has not yet been properly formulated,
the identification of various non-perturbative dualities has led
to a much deeper understanding of the big picture. One important
program is to characterize the moduli space of vacua as
completely as possible.
Superstring vacua are most easily understood when they have many
unbroken supersymmetries, but a realistic vacuum should have no
unbroken supersymmetries at all. A possible viewpoint is that we should work
our way towards the study of realistic vacua in small steps starting
from ones with a lot of supersymmetry. As the lessons at one stage are
learned, we can build on that experience at the next stage with the
number of supersymmetries cut in half. For a given number of unbroken
supersymmetries, the number of non-compact
space-time dimensions is also an issue, since the classification of
vacua becomes richer and more subtle as this number is decreased.

The maximum possible number of supersymmetries is 32, corresponding
to N=1 in eleven dimensions, N=2 in ten dimensions,
N=4 in six dimensions, or N=8 in four dimensions. The next case of interest
is 16 unbroken supersymmetries, corresponding to N=1 in ten dimensions,
N=2 in six dimensions, or N=4 in four dimensions. In all of these cases
we have a pretty good grasp of the complete moduli space of superstring
vacua. This is not to imply that everything about
these theories is understood, just that we can enumerate them
and identify their massless spectra in four or more dimensions.

It now seems timely to make a concerted effort to classify possible
vacua with 8 unbroken supersymmetries, corresponding to N=1 in
six dimensions or N=2 in four dimensions. This is certainly a
challenging problem and will take some time to sort out. For one
thing, the pioneering work of Seiberg and Witten \cite{sw94}
has taught us that
for N=2  models in four dimensions the quantum moduli space is
different from the classical one. Another indication that the
classification of such theories is a challenging problem is the fact that
many, but not all, are given by compactification of a Type II
theory on a Calabi--Yau manifold, and the classification of
Calabi--Yau manifolds is still far from complete.

A somewhat more modest problem is to classify N=1 vacua in six dimensions.
This problem should be more tractable for a number of reasons.
First, their number should be far fewer. Second, the possibilities
are significantly constrained by the requirements of anomaly cancellation.
To appreciate these two points, one should recall the situation
when there are 16 supersymmetries. In that case the maximum
dimension is ten, and the requirements of anomaly cancellation
imply that in ten dimensions
there are just two possibilities, corresponding to
gauge groups $SO(32)$ or $E_8 \times E_8$. Recall that when this
result was obtained \cite{gs84}, Type I superstrings were known,
but it was not clear whether there was a theory that realized
the $E_8 \times E_8$ group. This led to the discovery of the
heterotic string theory shortly thereafter. Ten-dimensional
N=1 models are completely specified at low energy by the choice of
the gauge group. In the case of N=1 models in six dimensions, on
the other hand, a complete characterization of the low energy
dynamics also requires specifying the representation of the
gauge group to which the massless hypermultiplets belong.

One obvious way to obtain N=1 models in six dimensions is to
compactify either of the two N=1 ten-dimensional models on a K3 manifold.
Aspects of this analysis have been discussed by a number of authors
\cite{gsw85,rsss85,walton88,erler94}. The generic result is described
in section 2 and the symmetry enhancement that is achieved at a
special (Gepner) point in the K3 moduli space is described in
section 3. All the models obtained in this way have a gauge group
with rank less than or equal to 20. They can be understood within
the framework of perturbative heterotic string theory using
standard conformal field theory technology.

In recent work, Witten has shown that shown that in the case
of the $SO(32)$ theory compactified on K3,
small instantons can give rise to non-perturbative symmetry enhancement
\cite{wittena}.
The largest gauge group that can be achieved in this way is
$SO(32) \times Sp(24)$, which has rank 40.
In section 4 we verify that this model
satisfies the requirements of anomaly cancellation. This
is a very non-trivial check. The anomaly analysis suggests that
it is very difficult to break the $SO(32)$ part of the gauge group,
and that nothing like this is going to work for $E_8 \times E_8$.
After the fact, it is evident that this model
could have been discovered by looking for new ways to
satisy the anomaly cancellation requirements. This lesson motivates
exploring whether they have other non-trivial solutions.
In section 5 we present two new classes of solutions for which the gauge
group is $SO(2n+8) \times Sp(n)$
or $SU(n) \times SU(n)$. It is quite surprising that gauge
groups of arbitrarily high rank can be consistent.
Even though realizations of these
solutions in string theory are not yet known, the
previous experience with $E_8 \times E_8$ in ten dimensions
suggests that it may be worthwhile looking for them.

\section{Review of Anomaly Cancellation Conditions}

$N=1$ supersymmetry in
six dimensions involves four types of massless multiplets.
Classifying massless particles by  representations of the little group $O(4)
\approx SU(2) \times SU(2)$, and labelling $SU(2)$ representations by their
multiplicities $(2J + 1)$, they are

$({i})$ gravity:  $({\bf 3},{\bf 3}) + 2 ({\bf 2,3}) + ({\bf 1,3})$

$({ii})$ tensor:  $({\bf 3,1}) + 2 ({\bf 2,1}) + ({\bf 1,1})$

$({iii})$ vector:  $({\bf 2,2}) + 2 ({\bf 1,2})$

$({iv})$ hyper:  $2({\bf 2,1}) + 4 ({\bf 1,1})$.

\noindent A general $N=1$ model has massless content given by
$(i) + n_T (ii) + n_V (iii)
+ n_H (iv)$.
With the exception of the final paragraph,
we will only consider the case $n_T = 1$, which is what
one expects for heterotic string compactifications.
The vector multiplets
belong to the adjoint representation of the gauge group $G$, and so $n_V =
{\rm dim}~G$.
The hypermultiplets belong to some representation $R$ of the group.
CPT invariance requires that $R$ is a real representation.\footnote{Models that
violate this condition were proposed in \cite{erler94}.}
{}From this it follows that
$n_H = {\rm dim}~R$.  (However, as explained in \cite{wittena},
if $R$ is pseudoreal, it can be realized by ${1\over 2}
{\rm dim}~R$ hypermultiplets.)  The knowledge of $G$ and $R$ completely
characterizes the low
energy dynamics and goes a long way towards characterizing the associated
string theory dynamics.

The requirement of cancellation of all gauge and
gravitational anomalies is a stringent condition on the possible choices of $G$
and $R$.  The anomalies
are characterized by a formal 8-form (a characteristic class)
made from the curvature and gauge field 2-forms.  One requirement is the
cancellation of the $\tr R^4$ term, where $R$ is the curvature 2-form.  This
leads to the requirement
\begin{equation}
n_H = n_V + 244.
\label{nheqn}
\end{equation}
The general condition for cancellation of the remaining anomalies has been
given previously \cite{gsw85,erler94}.
To keep things relatively simple, the result will be
given for the special case that $G = \otimes G_\alpha$ is semi-simple,
{\it i.e.}, there are no $U(1)$ factors. Assuming that eq. (\ref{nheqn})
is satisfied, and normalizing the remaining anomaly 8-form so
that the coefficient of $(\tr R^2)^2$ is unity, gives
\begin{equation}
I = (\tr R^2)^2 + {1\over 6} \tr R^2 \sum_\alpha X_\alpha^{(2)} - {2\over 3}
\sum_\alpha X_\alpha^{(4)} + 4 \sum_{\alpha < \beta} Y_{\alpha\beta} ,
\label{itotal}
\end{equation}
where
\begin{equation}
X_\alpha^{(n)} = \Tr F_\alpha^n - \sum_i n_i \tr_i F_\alpha^n
\label{xdef}
\end{equation}
\begin{equation}
Y_{\alpha\beta} = \sum_{ij} n_{ij} \, \tr_i F_\alpha^2 \, \tr_j F_\beta^2.
\label{ydef}
\end{equation}
The notation is as follows:  The symbol $\Tr$ denotes a trace in the adjoint
representation and $\tr_i$ denotes a trace in the representation $R_i$ (of the
simple group $G_\alpha$).  $n_i$ is the number of hypermultiplets in the
representation $R_i$ of $G_\alpha$ and $n_{ij}$ is the number in the
representation $(R_i, R_j)$ of $G_\alpha \times G_\beta$.  These numbers are
usually positive integers, but can be half-integral when pseudo-real
representations occur.  The cancellation of the remaining gravitational, gauge,
and mixed anomalies by the mechanism in \cite{gs84} requires
that the anomaly 8-form should factorize as
\begin{equation}
I = (\tr R^2 + \sum_{\alpha} u_\alpha \tr F_\alpha^2)
(\tr R^2 + \sum_{\alpha} v_\alpha \tr F_\alpha^2), \label{ifactor}
\end{equation}
where $u_{\alpha}$ and $v_{\alpha}$ are numerical coefficients and
$\tr F_\alpha^2$ is evaluated in a convenient (``fundamental'')
representation of $G_\alpha$.

Given a consistent model solving the anomaly cancellation conditions,
one can trivially obtain other ones by enlarging the gauge group
by an arbitrary gauge group $G'$ and adding hypermultiplets in the
adjoint representation of $G'$. Models with this structure are
{\it reducible} and will not be considered.

The best way to understand the meaning of eqs. (\ref{itotal}--\ref{ifactor})
is by means of a
few examples.  The standard examples, which follow, correspond to
compactification of the $SO(32)$ and $E_8 \times E_8$ heterotic strings on a
generic $K3$.  In each case, one identifies the spin connection (which belongs
to $SU(2)$ in the case of $K3$) with a suitable $SU(2)$ subgroup of the gauge
group.  This means that one chooses a background $SU(2)$ gauge bundle on the
$K3$ with instanton number 24.  Decomposing $SO(32)$ as $SO(28) \times SO(4)$
and identifying the spin connection with one of the two
$SU(2)$'s in $SO(4) = SU(2)
\times SU(2)$, leaves an unbroken gauge group $G = G_1 \times G_2$ with $G_1 =
SO(28)$ and $G_2 = SU(2)$.  Then the multiplicities of various hypermultiplet
representations can be read off from standard index-theorem formulas given in
\cite{gsw85}.  One finds
\begin{equation}
R = 10 ({\bf 28,2}) + 65 ({\bf 1,1}),
\end{equation}
where 20 of the singlets come from the gravitational sector and 45 from the
matter sector of the ten-dimensional theory.   The first term in $R$ is better
thought of as 20 copies of ${1\over 2} ({\bf 28,2})$, since $({\bf 28,2})$ is
pseudoreal.  Note that altogether $n_V = 378+3 = 381$ and $n_H = 560 + 65 =
625$, which satisfies eq. (\ref{nheqn}).

The construction ensures that the anomaly
cancellation conditions are satisfied, but let's verify them anyway.  This
requires the identities
\begin{equation}
\Tr F^4 = (n - 8) \tr F^4 + 3 (\tr F^2)^2
\end{equation}
\begin{equation}
\Tr F^2 = (n-2) \tr F^2
\end{equation}
for $SO(n)$ and
\begin{equation}
\Tr F^4 = 8 (\tr F^2)^2
\end{equation}
\begin{equation}
\tr F^4 = {1\over 2} (\tr F^2)^2
\end{equation}
\begin{equation}
\Tr F^2 = 4 \tr F^2
\end{equation}
for $SU(2)$.  Using these in eqs. (\ref{xdef}) and (\ref{ydef}), we obtain
\begin{equation}
X_1^{(2)} = 6 \tr F_1^2 , \quad X_1^{(4)} = 3 (\tr F_1^2)^2
\end{equation}
\begin{equation}
X_2^{(2)} = - 276 \tr F_2^2, \quad  X_2^{(4)} = - 132 (\tr F_2^2)^2
\end{equation}
\begin{equation}
Y_{12} = 10 \tr F_1^2 \tr F_2^2.
\end{equation}
The important point is that all $\tr F^4$ terms cancel and the remaining
expression factorizes as follows
\begin{equation}
I = (\tr R^2 - \tr F_1^2 - 2 \tr F_2^2) (\tr R^2 + 2 \tr F_1^2 - 44 \tr F_2^2).
\end{equation}

The analysis of the $E_8 \times E_8$ model is quite similar.  One can identify
the spin connection with the $SU(2)$ factor in
the decomposition $E_8 \supset E_7 \times SU(2)$ of one of the two $E_8$'s,
leaving an unbroken gauge group $G = G_1 \times G_2$, where $G_1 = E_8$ and
$G_2 = E_7$.  The massless hypermultiplets are
\begin{equation}
R = 10 ({\bf 1,56}) + 65 ({\bf 1,1}).
\end{equation}
Only singlets of $E_8$ occur, which is therefore a ``hidden sector'' group.
As before, $n_V = 381$ and $n_H = 625$.  To analyze the anomaly formula, we
need the identities
\begin{equation}
\Tr F^4 = {1\over 100} (\Tr F^2)^2
\end{equation}
for $E_8$ and
\begin{equation}
\Tr F^4 = {1\over 6} (\tr F^2)^2
\end{equation}
\begin{equation}
\tr F^4 = {1\over 24} (\tr F^2)^2
\end{equation}
\begin{equation}
\Tr F^2 = 3 \tr F^2
\end{equation}
for $E_7$.  Here $\tr$ is evaluated in the ${\bf 56}$.  Using these, one
obtains a factorized anomaly
\begin{equation}
I = (\tr R^2 - {1\over 30} \Tr F_1^2 - {1\over 6} \tr F_2^2)
(\tr R^2 + {1\over 5} \Tr F_1^2 - \tr F_2^2). \label{e8e7}
\end{equation}

When one has anomaly-free models,
such as the two given above, one can obtain many more
by Higgsing.  The unique way vector and hypermultiplets can become
massive is for one of each to pair up to give a massive vector multiplet.  Note
that this preserves $n_H - n_V = 244$.  There is a triplet of $D$ terms,
quadratic in the hypermultiplet scalar fields, for each  generator of $G$.  The
$4n_H$ scalars can be given arbitrary vevs that maintain the vanishing of all
$3n_V$ $D$ terms.  Generically, this breaks much of the gauge symmetry giving
many more consistent solutions of the anomaly equations.  In this way one
probes different phases of a connected moduli space of vacua.

Let us consider examples of such Higgsing for the two models we have presented.
Starting with $SO(28) \times SU(2)$, we can break it to $G = SO(N)$, with
$N\leq 28$.  The hypermultiplet representation then consists of $N-8$ copies of
the {\bf N} representation
and the number of singlets required to maintain $n_H - n_V = 244$.  For all
these models the anomaly factorizes in the form
\begin{equation}
I = (\tr R^2 - \tr F^2) (\tr R^2 + 2 \tr F^2).
\end{equation}

The $E_8 \times E_7$ model can be Higgsed in a similar manner.  Before
describing that, let us consider the more general problem of $G = E_8 \times
G'$, where $G'$ is an arbitrary semi-simple group and all hypermultiplets are
$E_8$ singlets.  In this case (assuming eq. (\ref{nheqn})) one has
\begin{equation}
I = (\tr R^2)^2 + {1\over 6} \tr R^2 \Tr F_1^2 -
{1\over 150} (\Tr F_1^2)^2 + A \tr R^2 + B, \label{abeqn}
\end{equation}
where $A$ and $B$ depend on $G'$.  This can factorize if and only if
\begin{equation}
B = {6\over 49} A^2, \label{bform}
\end{equation}
in which case we obtain
\begin{equation}
I = (\tr R^2 - {1\over 30} \Tr F_1^2 + {1\over 7} A) (\tr R^2 + {1\over 5} \Tr
F_1^2 + {6\over 7} A).
\end{equation}
Equation (\ref{e8e7}) is of this form.

Now consider Higgsing the $E_8 \times E_7$ model to obtain $E_8 \times E_6$.
The $E_6$ representations that describe the hypermultiplets turn out to be  $9
\cdot {\bf 27} + 9 \cdot \overline{{\bf 27}} + 84 \cdot {\bf 1}$.  One finds
that $A = - {7\over 3} \tr F_2^2$ and
$B = {2\over 3} (\tr F_2^2)^2$, so that eq.
(\ref{bform}) is satisfied.  Further Higgsing to $E_8 \times SO(10)$ gives
hypermultiplets $18 \cdot {\bf 10} + 8 \cdot {\bf 16} + 8 \cdot \overline{{\bf
16}} + 101 \cdot {\bf 1}$.  This time
$A = - 7 \tr F_2^2$ and $B = 6 (\tr F_2^2)^2$,
which also satisfies eq. (\ref{bform}).  This sequence of models
corresponds to the ones considered in \cite{kachru}.

\section{Symmetry Enhancement}

In the preceding example we gave ``standard'' K3 compactification models and
showed how ones with less symmetry can be reached by Higgsing.
One can also find models with more gauge symmetry.  These correspond to special
subclasses of K3's (or orbifolds) that give symmetry enhancement -- the
so-called Gepner points.  One could derive these from first principles or (more
simply) guess the result by playing around with the anomaly formulas.  I will
choose the latter route.  It will suffice to find the example with maximal
symmetry enhancement, since any others, including those in the
preceding section, can then be obtained by Higgsing.

I claim that the maximal symmetry that be obtained in this way is
\begin{equation}
E_8 \times E_7 \times [SU(2)]^5 ~~~{\rm or}~~~ SO (28) \times [SU(2)]^6.
\end{equation}
Both of these models have rank 20, which is the most that can be accommodated
in  a perturbative conformal field theory treatment of the heterotic string
(for $N=1$ and $D=6$).
The possibility of a $[U(1)]^5$ symmetry enhancement appears in the literature,
but (as far as I am aware) $[SU(2)]^5$ is new.
The $E_8 \times E_7 \times [SU(2)]^5$ example, which has $n_V = 396$ and hence
requires $n_H = 640$, works as follows.  The hypermultiplet representation
(suppressing the $E_8$ label, which is always a singlet) is
\begin{equation}
R = [({\bf 56, 2, 1,1,1,1})
+ ({\bf 1; 2,2,2,2,1})] + 4 \ {\rm perms}.
\end{equation}
Note that this gives $n_H = 640$ without the need for any singlets.  This is
what one might expect for a maximally symmetric Gepner point.  The quantities
$A$ and $B$ in eq. (\ref{abeqn}) in this case turn out to be
\begin{eqnarray}
A &=& {1\over 6} \left(-7 \tr F_1^2 - 84 \sum_{i = 1}^5 \tr
F_{2i}^2\right)\nonumber\\
B &=& - {2\over 3} \left(-{1\over 4} (\tr F_1^2)^2 - 36 \sum_{i = 1}^5 (\tr
F_{2i}^2)^2\right)\nonumber\\
&+& 4 \left( \tr F_1^2 \sum_{i=1}^5 \tr F_{2i}^2 +
12 \sum_{i<j} \tr F_{2i}^2 \tr F_{2j}^2\right).
\end{eqnarray}
These satisfy $B = {6\over 49} A^2$ as required.

It appears that all $N=1 $ $D = 6$ models that can be understood within the
framework of perturbative heterotic string theory can be obtained by suitable
Higgsing of these two models.  So at this point we have two disconnected
components for the $D = 6$ slice of the moduli space of vacua with
eight unbroken supersymmetries ({\it i.e.}, $N=1$).  But,
as we will see, there is more.

\section{Small Instantons}

In a recent paper, Witten showed that in the case of the $SO(32)$ theory there
are non-perturbative possibilities for symmetry enhancement associated with
instantons of vanishing size \cite{wittena}.
They correspond to Dirichlet
five-branes, which carry additional symplectic group symmetry.  Since the total
instanton number should be 24, the maximal possibility for symmetry enhancement
is to have 24 coincident five-branes carrying an $Sp (24)$ gauge symmetry.  The
notation is that $Sp(k)$ refers to a compact group of rank $k$, which is
sometimes called $USp (2k)$ by other authors (including myself on occasion).
Its fundamental representation
has dimension $2k$ and the adjoint has dimension $k(2k + 1)$.  An antisymmetric
tensor of dimension $k(2k - 1)$ is reducible into a singlet plus the rest.  The
most symmetric small instanton model has $G = G_1 \times G_2$ with $G_1 =
SO(32)$ and $G_2 = Sp (24)$.  This has rank 40 and dimension $n_V = 1672$.  Its
existence cannot be understood in terms of conformal field theory, since the
small instantons are inherently non-perturbative, as explained
in \cite{wittena}.  Our
purpose here is to examine anomaly cancellation for this model.

The number of hypermultiplets must be $n_H = n_V + 244 = 1916$.  From the
analysis in \cite{wittena},
we know that there is ${1\over 2} ({\bf 32,48})$ corresponding to open
strings with one end attached to the five-brane.  Note that the factor of
${1\over 2}$ is allowed here, because ${\bf 48}$ is pseudoreal.  In addition,
there is an antisymmetric tensor representation corresponding to open strings
with both ends attached to the five-brane.  This gives $({\bf 1,1127}) + ({\bf
1,1})$.  We now have 1896 hypermultiplets.  Thus, there must
also be 20 singlets, which are just the usual gravitational moduli for $K3$
compactification.

Now the anomaly analysis can be carried out using the $Sp(k)$ identities
\begin{eqnarray}
\Tr F^4 &=& (2k + 8) \tr F^4 + 3 (\tr F^2)^2\nonumber\\
\Tr F^2 &=& (2k + 2) \tr F^2\nonumber\\
\tr_A F^4 &=& (2k - 8) \tr F^4 + 3 (\tr F^2)^2\nonumber\\
\tr_A F^2 &=& (2k - 2) \tr F^2,
\end{eqnarray}
where $A$ refers to the antisymmetric tensor representation.  Using these
formulas we have
\begin{eqnarray}
X_1^{(2)} &=& 6 \tr F_1^2 ~,\quad \quad X_1^{(4)} = 3 (\tr F_1^2)^2\nonumber\\
X_2^{(2)} &=& - 12 \tr F_2^2 ~, \quad X_2^{(4)} = 0\nonumber\\
Y_{12} &=& {1\over 2} \tr F_1^2 \tr F_2^2. \label{xandy}
\end{eqnarray}
Substitution into the anomaly formula gives the factorized expression
\begin{equation}
I = (\tr R^2 - \tr F_1^2) (\tr R^2 + 2 \tr F_1^2 - 2 \tr F_2^2).
\label{smallfact}
\end{equation}
This is an impressive confirmation of Witten's result.  In particular, the
cancellation of the $\tr F^4$ terms for the $Sp$ factor required the $SO(m)$
factor to be $SO(32)$ and the cancellation of the $\tr F^4$ terms for the $SO$
factor required the $Sp(n)$ factor to be
$Sp(24)$ -- no other $m$ or $n$ would work
with these representations.

Let us now examine what other models can be reached by Higgsing this model.  It
is easy to see that the $Sp (24)$ factor can be broken to a subgroup
$\otimes_{i=1}^n Sp (k_i)$ with $\sum k_i = 24$.  In this case the ${1\over 2}
({\bf 32,48})$ decomposes in the obvious way without any of these
hypermultiplets being
eaten.  All the hypermultiplets that are eaten
come from the antisymmetric tensor, leaving a
sum of antisymmetric tensors (including the singlets) for each of the $Sp(k_i)$
factors.  Also, the 20 singlets of gravitational origin survive untouched.  The
anomaly analysis works as above with $tr F_2^2 \rightarrow \sum_{i=1}^n tr
F_{2i}^2$.  This result is exactly as expected based on the analysis
of \cite{wittena}.

The $SO(32)\times Sp(24)$ model can be generalized to the case of $n\leq 24$
coincident small instantons. This requires embedding $24-n$ units of
instanton number in the $SO(32)$. The resulting gauge group is
$SO(8+n) \times Sp(n)$.\footnote{I am grateful to E. Witten for bringing this
possibility to my attention.} In this case the hypermultiplets
consist of ${1\over 2} ({\bf 8+n,2n}) + {24-n \over 2} ({\bf 1,2n})$,
as well as the antisymmetric tensor representation of $Sp(n)$. The
number of singlets (besides the one associated with the antisymmetric tensor)
is $20 + {1\over2} (n-24)(n-21)$. The anomaly analysis works as before,
and the result is again given by eqs. (\ref{xandy}) and ({\ref{smallfact}).
The $Sp(n)$ can again be broken by Higgsing, as described in the
preceding paragraph.

Another interesting question is whether small instantons can be accomodated in
$E_8 \times E_8$ models.  $SO(32)$ and $E_8 \times E_8$ models belong to a
common moduli space after compactification of each of them on a circle.
However, this is not the case for $K3$ compactification.  This
is the same as for type IIA and IIB theories, which join up after $S^1$
compactification, but remain distinct upon $K3$ compactification.  This means
that $E_8 \times E_8$ models on $K3$ do not have a dual type I description, and
D-branes are meaningless for them.  Wisely, ref. \cite{wittena}
made no claims for small
instanton effects in $E_8 \times E_8$ models.  From the point of view of
anomaly equations, it is obvious that there are no solutions of the form $E_8
\times E_8 \times G'$, at least if one assumes that all hypermultiplets are
singlets of both $E_8$'s. Non-singlet representations quickly lead to very
large
values of $n_H$, which appear unlikely to lead to any consistent new
possibilities.

\section{New Anomaly-Free Models}

Maybe there are six-dimensional $N=1$ string vacua that arise
from other non-perturbative mechanisms. One way to identify candidates
is to find new solutions of the anomaly cancellation conditions.
A non-systematic search turned up
two new classes of solutions of the anomaly conditions, which may be
of some interest.  For the first class, the gauge group is
\begin{equation}
G = SO(2n + 8) \times Sp (n),
\end{equation}
which has rank $2n + 4$.  Since $n$ is an arbitrary non-negative integer, the
rank may be arbitrarily large. The hypermultiplet content of these models is
given by
\begin{equation}
R = ({\bf 2n + 8,2n}) + 272 ({\bf 1,1}).
\end{equation}
Since ${\bf 2n}$ is pseudoreal, the first term is really two
copies of ${1\over 2}
({\bf 2 n} + {\bf 8,2n})$.  The number of singlets is determined from the
requirement $n_H = n_V + 244$.  Despite the superficial resemblance to the
examples in the preceding section,
there are two important differences.  One is the factor of two
mentioned above and the second is the absence of an antisymmetric tensor
representation of the symplectic group.  The anomaly analysis is easy using the
formulas in the preceding sections.  One finds the factorized result
\begin{equation}
I = (\tr R^2 - \tr F_1^2 + \tr F_2^2) (\tr R^2 + 2 \tr F_1^2 - 2\tr F_2^2),
\end{equation}
for all values of $n$.  This result depends on a number of ``miracles,'' so I
expect it to have physical significance.
The most straightforward Higgsing of these models simply
decreases the value of $n$. This means that these models form a single
connected
structure.  Roughly speaking, there ought to be an infinite-dimensional group
that underlies all of them.

The second class of models has
\begin{equation}
G = SU(n) \times SU(n)
\end{equation}
with hypermultiplets in the representation
\begin{equation}
R=({\bf n, \bar n}) + ({\bf \bar n, n}) + 242 ({\bf 1, 1}).
\end{equation}
Using the $SU(n)$ formulas (and $\tr_n = \tr_{\bar n} = \tr)$
\begin{eqnarray}
\Tr F^4 &=& 2n \tr F^4 + 6 (\tr F^2)^2\nonumber\\
\Tr F^2 &=& 2n \tr F^2,
\end{eqnarray}
one finds that anomaly factorizes as follows
\begin{equation}
I = (\tr R^2 - 2 \tr F_1^2 + 2 \tr F_2^2)
(\tr R^2 + 2 \tr F_1^2 - 2 \tr F_2^2).
\end{equation}
Thus, we have a second infinite family of models with unbounded rank.
An intriguing feature of this class of models, not shared by the first
one, is that the corresponding $N=2$ four-dimensional gauge theory
is superconformal (or finite).

The two classes of models do share another interesting feature.
The gauge groups are the bosonic subgroups of simple Lie superalgebras
$OSp (2n +8|n)$ and $SU(n|n)$.
Moreover, the non-singlet
hypermultiplets are in correspondence with the odd elements of the
superalgebras. These properties seem closely related to the
observation in ref. \cite{harvey} that the BPS states of certain
$N=2$ $D=4$ models have a similar relationship to infinite superalgebras.
It is also intriguing that the factorized anomalies only involve the
`supertrace' combinations $\tr F_1^2 - \tr F_2^2$.
It is tempting to conjecture that the $OSp (2n +8|n)$ class
of models has an explanation in terms of unoriented open strings
and the $SU(n|n)$ class
of models has an explanation in terms of oriented open strings.

A. Dabholkar has pointed out that if one adds eight tensor multiplets
to these models (for a total of nine),
then eq. (1) is replaced by $n_H - n_V =12$
and the $(\tr R^2)^2$ term in the anomaly cancels. In this case
the number of singlets becomes 40 in the $OSp(2n+8 | n)$ models
and 10 in the $SU(n|n)$ models. Also, the absence of the $(\tr R^2)^2$
term implies that the gauge invariant field strength $H$ (which
describes the self-dual tensor of the gravity multiplet and one of
anti-self-dual matter tensors) contains a Yang-Mills Chern--Simons
term and no Lorentz Chern--Simons term.

\section{Acknowledgments}

I am grateful to R. Leigh and N. Seiberg for helpful discussions of the
analysis in section 4, to A. Dabholkar and J. Gauntlett for
discussions concerning possible string theory interpretations of the models
in section 5, and to I. Bars for a discussion of supergroups.

\vfill\eject

\end{document}